# Collaboration Encouraging Quantum Secret Sharing Scheme with Seal Property

Xiaogang CHENG[1], Ren GUO[2]

1. College of Computer Science and Technology, Huaqiao University, Xiamen 361021, China
2. College of Business Administration, Huaqiao University, Quanzhou 362021, China

**Abstract**: A new concept of quantum secret sharing is introduced, in which collaboration among participants are encourage. And the dealer can ask the participants to send back their share and revoke the secret before a predefined date or event, i.e. so-called seal property. We also give two concrete constructions of CE-QSS-Seal (Collaboration-Encouraging Quantum Secret Sharing with Seal property) scheme. The first one is unconditional secure and achieve the optimal bound of a seal scheme. The second one improve the optimal bound of seal by introducing post-quantum secure computational assumption.

**Keywords**: quantum secret sharing, collaboration encouraging, quantum seal, post-quantum cryptography, certified deletion

**1.Introduction**

In SS (Secret Sharing) scheme, a dealer can distribute a secret to several parties, and only some predefined subset (the so-called access structure) of the parties together can recover the secret [1,2,3,4]. There are many extensions of SS. In VSS (Verifiable SS) [5], the legitimacy of the secret and the shares can be verified. The shares in PSS (Proactive SS) [6,7] can be updated periodically to enhance the security of the secret over a long period of time. In RSS (Rational SS) [8,9], the participants are assumed to be rational rather than honest of malicious assumptions in usual cryptographic settings. With the advent of quantum information, recently there are many research on QSS (Quantum SS) [10-15], in which the secret and the shares can be quantum particles.

In certain SS scenarios, suppose the dealer want the participants to reconstruct the secret only when he gives the order or some predefined event happens, or on a predefined date. Before that, the dealer can check the honesty of the participants by asking them to send back their intact share to prove that they have not try to reconstruct the secret. This strategy is simply impossible in classical setting, since classical share can be easily copies any number of times. While in quantum setting, there is the famous fundamental non-cloning theorem which prevent copying unknown quantum state. In this paper, we show how to implement this idea quantum mechanically. This concept is related to quantum seal [16-25], in which there is only one participant for sharing the secret.

Another property of our novel quantum secret sharing concept is CE (Collaboration-Encouraging), which means that more parties participate in the reconstruction process, the easier the secret can be recovered. So as to encourage collaboration among the participants. We call our novel secret sharing scheme CE-QSS-Seal (Collaboration-Encouraging Quantum Secret Sharing with Seal property).

We also give two concrete CE-QSS-Seal constructions. The first CE-QSS-Seal scheme we constructed achieve the optimal bound of quantum seal [25]. Namely the secret can be constructed with 100% probability, and the cheating behavior of the participants can be caught with probability of 50%. To improve this scheme, we then use post quantum cryptography [26] and PKE (Public

Key Encryption) with certified deletion [27,28] to construct a nearly perfect scheme, i.e., the probability of reconstructing the secret is 100%, and also the cheating behavior of the participants can be caught with probability of almost 100%.

The paper is arranged as follows. In Section 2, we give the definition of CE-QSS-Seal scheme and introduce some tools used in the following construction. Then concrete constructions of CE-QSS-Seal scheme are given in Section 3. The security of our construction is analyzed in section 4. And finally, we conclude in Section 5.

**2.Preliminaries**

**Definition 1.** CE-QSS-Seal scheme works as following.

There are $n$ participants for sharing a secret. The dealer can break the secret (quantum or classical) into $n$ shares (quantum or classical), and send each share to a participant. A predefined subset of the $n$ participants can together reconstruct the secret. While a non-legitimate subset cannot reconstruct the secret. It is called a perfect secret sharing scheme if a non-legitimate subset cannot gain any information about the secret.

In the usual $(t,n)$ threshold secret sharing scheme, the secret can be reconstructed only if there are more than $t$ participants present. After the threshold $t$, more parties will not bring any benefit to the reconstruction process. CE property means that collaboration among participants is encouraged, i.e. the secret can be easier to reconstruct if there are more parties are involved in the reconstruction process.

For the Seal property, the dealer can set a predefined date/time or event, before which the secret should not be reconstructed. And the dealer can check the honesty of the participants by asking them to return their share before the predefined date or event to prove that they have not try to reconstruct the secret. This is simply impossible in classical information, since classical share can be copied easily any number of times. Hence the share should be quantum. In Fig.1, we sketch QSS with seal property. The dealer sends quantum shares to each participant. And all the participants can come together to recover the secret. While the dealer can also ask all the participants to send back their quantum shares before a predefined date/event (dotted line in the figure).

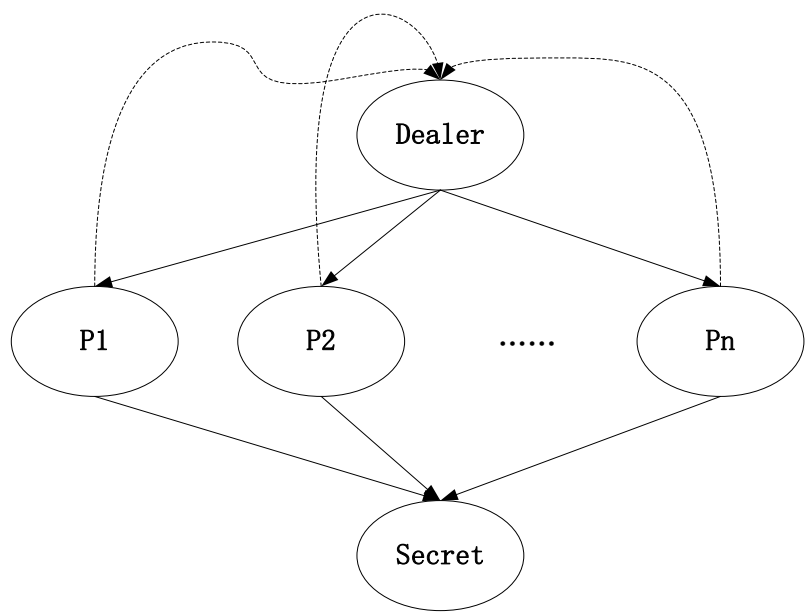

Fig. 1. Quantum secret sharing scheme with seal property

For our CE-QSS-Seal scheme, there are two security properties:

1. CE-QSS property: The successful reconstruction probability should be diminishing exponentially with the number of missing parties. Thus, encourage collaboration among the participants.

2. Seal property: The dealer asks the participants to send back the quantum shares before a predefined date or event for checking the honesty of the parties. If some parties are dishonest and try to reconstruct the secret beforehand. The dealer can find out this dishonest behavior with high probability. And the successful reconstruction probability should also be high if all the parties are honest and the predefined date or event has come.

In the following construction, we use a cryptographical primitive called PKE (Public Key Encryption) with Certified Deletion:

To encrypt a bit $b$, randomly generate n-bit long classical number $x$ and $\theta$, and prepare the following quantum state:

$$|x\rangle_\theta = \otimes_i |x_i\rangle_{\theta_i}$$

In which the qubit $|x_i\rangle$ is encoded with computational basis $\{|0\rangle, |1\rangle\}$ if $\theta_i = 0$, and encoded with Hadamard basis $\{|+\rangle, |-\rangle\}$ if $\theta_i = 1$.

Then using post-quantum secure encrypt scheme to encrypt $\theta$ and the bit b with mask $\oplus_{i:\theta_i=0} x_i$. And the final ciphertext is:

$$|x\rangle_\theta, ENC(\theta, b \oplus \bigoplus_{i:\theta_i=0} x_i)$$

The receiver of this ciphertext can prove that he has not try to decrypt it by measuring the quantum state $|x\rangle_\theta$ by Hadamard basis, and send the measurement result to the sender of the ciphertext. Of course, the sender can easily check the honest of the receiver with the knowledge $x$ and $\theta$. And also note that after this Hadamard measurement by the receiver, the encrypted bit $b$ will be hided completely from the receiver. Since measuring the quantum state by the Hadamard basis completely destroy the information about those $x_i$, for which the corresponding $\theta_i$ is 0 and encoded with computational basis. Then b is hidden information-theoretically from the receiver even though he can decrypt the classical part the ciphertext

$$ENC(\theta, b \oplus \bigoplus_{i:\theta_i=0} x_i)$$

To get $\theta$ and $b \oplus \oplus_{i:\theta_i=0} x_i$.

### 3. Construction of CE-QSS-Seal scheme
### 3.1. Unconditionally secure CE-QSS-Seal with optimal seal bound

The dealer generates a DHZ-like $n$-qubit long quantum entangled state as following:

$$|x\rangle + |\bar{x}\rangle$$

$\bar{x}$ is simply the bit-wise negation of $x$. Such as:

$$|1001111\ldots 0\rangle + |0110000\ldots 1\rangle$$

Then send each qubit to the $n$ parties respectively as share of our CE-QSS-Seal scheme. When encoding the secret, the dealer and all the parties agree that the messages represented by $x$ and $\bar{x}$ are the same, i.e. we only use half of the $n$-bit long space for coding.

If all parties are present, then the reconstruction process is easy. All the parties put their quantum particles together and measure in the computational basis, then one of the classical bit patterns $\{x, \bar{x}\}$ should be gotten. And this is the secret. Since by the convention mentioned above, both $x$ and $\bar{x}$ represent the same secret.

Now suppose one party is missing. Still by measuring the $n-1$ particles by the computational basis, all but one bit of $x$ or $\bar{x}$ will be gotten. Hence by randomly guessing the remaining bit, the correct secret will be gotten with probability of $1/2$.

If two parties are missing, by randomly guessing the remaining two bits, the correct secret can be gotten with probability of 1/4.

In general, when $k$ parties are missing, the correct secret can be gotten with probability $1/2^k$, by randomly guessing the missing $k$ bits.

Hence if there are few parties are missing in the reconstruction process, the secret can be successfully reconstructed with high probability. While there are many parties are missing, the successful probability of reconstructing the secret is very small. I.e., the probability decreased exponentially with the number of missing parties.

Before the specified date or event, the participated parties are not supposed to reconstruct the secret. Classically it is impossible for the dealer to check if the parties are honest or not, since classical shares can be copied easily. While in our quantum SS, the deal can check the honesty of the parties by asking all the participants to send back their quantum particles. If all the parties are honest, then the dealer should get the original GHZ-like state:

$$|x\rangle + |\bar{x}\rangle$$

Then measure this pure state in the quantum basis:

$$|x\rangle + |\bar{x}\rangle, |x\rangle - |\bar{x}\rangle$$

The measurement result should always be the first vector. While if one or more dishonest parties try to reconstruct the secret beforehand. Then the dealer will get the state $|x\rangle$ or $|\bar{x}\rangle$, which will produce the two basis-vector will equal probability when measuring in the quantum basis mentioned above. So, with probability 1/2, the cheating behavior will be detected by the dealer.

Note that this 1/2 match the optimal bound in [25]. So, from information-theoretical security point of view, it is impossible to improve this probability. Next, we show how to improve this bound by post-quantum cryptography.

Here is a simple 10-parties example:

$$|0011100101\rangle + |1100011010\rangle$$

If all parties are present, and this quantum state is measured by the computational basis, we can get $0011100101$ or its bit-wise negation $1100011010$ with equal probability. As mentioned above, both are perfect good secret based on the encoding scheme. While if one party is missing (WLOG suppose the last party is missing), by measuring in the computational basis the remaining parties will get $001110010$ or $110001101$ with equal probability. Note that this is a 9-bit long number, and the last bit is completely unknown to the nine parties. By random guessing, they will get the correct secret with probability 1/2. By similarly analysis, the probability is 1/4 if two parties are missing and $1/2^k$ if $k$ parties are missing.

If some parties are dishonest and try to get the secret before the predefined date or event, they will measure their quantum particles by the computational basis, and this measurement will collapse the whole quantum state to $|0011100101\rangle$ or $|1100011010\rangle$ with equal probability. If at some point the dealer ask all the participants to return their shares and measure by the two-output basis:

$$|0011100101\rangle + |1100011010\rangle$$
$$|0011100101\rangle - |1100011010\rangle$$

If the measurement result is the first one, the dealer believe that the participants are honest. While if the measurement result is the second one, the dealer can be sure that some parties are dishonest and can punish them accordingly. It is obvious that the probability of catching the dishonest behavior is 1/2, and with 1/2 probability the dishonest behavior can escape the dealer's checking.

**3.2 Construction based on PKE with certified deletion**

1. The share of each party, is just the Public-Key encryption with Certified Deletion of a random bit $b$:

$$|x\rangle_\theta, ENC(\theta, b \oplus \bigoplus_{i:\, \theta_i=0} x_i)$$

2. Reconstruct the secret: each party just measure the quantum state $|x\rangle_\theta$ in the computational basis, then decrypt the encryption to get $\theta$ and $b \oplus (\bigoplus_{i:\, \theta_i=0} x_i)$. Then it is easy to unmask to get $b$ by XOR those bits of $x_i$ where $\theta_i$ is zero.

Finally, all the parties put their unmasked $b_i$ together, the final secret will be gotten:

$$b_1 b_2 b_3 \ldots b_n$$

3. If the dealer wants to retreat the secret beforehand for checking the honesty of the participants. He can ask all the parties to measure the quantum state $|x\rangle_\theta$ in the Hadamard basis, and send back the measurement result. The dealer can easily check if the participant is honest or not by checking the measurement result, in which the result bit should match with the corresponding bit of $x$ when the bit of $\theta$ is 1. Since in this case, the bit of $x$ is encoding with Hadamard basis. And this operation also erases the bits of $x$ where the corresponding bits of $\theta$ is 0, thus the share $b$ is erased permanently.

4. The analysis of Collaboration-Encouraging property is similar with above. If one party is missing, then one secret bit is missing and the correct secret can be guessed with probability of 1/2. If $k$ parties are missing, then the probability of successful guessing of the secret is $1/2^k$ as mentioned above.

5. For the seal property, this scheme is improved significantly compared with the first construction. Since now if a party is dishonest and try to get the secret before the specified date or event, he has to measure the quantum state $|x\rangle_\theta$ by the computational basis. Then when the dealer asks him to return the measurement result by the Hadamard basis, his dishonest behavior will be caught with very high probability, close to 1. Since all the qubit encoded with Hadamard basis has been destroy when he made the computational basis measurement.

Still another advantage is that now the dishonest behavior of each party can be traced. I.e., the dealer can find out which party (or parties) is (are) dishonest. Since he can check the Hadamard basis measurement result of each party. While in our first construction, the dealer can only find that some parties are dishonest, but he cannot find out which one(s). Since no matter how many parties measure the quantum state $|x\rangle + |\bar{x}\rangle$ in the computational basis, it will collapse to $|x\rangle$ or $|\bar{x}\rangle$. Hence the dealer cannot find out who are the dishonest one(s).

6. We can easily extend this construction based on PKE with certified deletion to a $(t, n)$ revocable threshold QSS scheme. Just encrypt each share $S_i$ of the Shamir's secret share scheme with PKE with certified deletion:

$$|x\rangle_\theta, ENC(\theta, S_i\ masked\ by\ x)$$

Where $x$ is a multi-qudit number and encode with computational or Fourier basis according to $\theta_i = 0$ or $\theta_i = 1$ respectively. And the mask of the share is the qudits corresponding to $\theta_i = 0$, while the qudits of x corresponding to $\theta_i = 1$ can be used for generate certificate of deletion similar with above. Now the XOR operation should be replaced by modular operations, i.e.,

$$S_i + \sum_{i:\theta_i=0} x_i \; mod \; d$$

Where $d$ is the dimension of the quantum qudit. For example, if $d = 6$, we can use the computational basis $\{|0\rangle, |1\rangle, |2\rangle, |3\rangle, |4\rangle, |5\rangle\}$. And the Hadamard basis (i.e., Fourier basis) is:

$$\{|\hat{0}\rangle, |\hat{1}\rangle, |\hat{2}\rangle, |\hat{3}\rangle, |\hat{4}\rangle, |\hat{5}\rangle\}$$

Where：

$$|\hat{i}\rangle = \sum_j \omega^{ij} |j\rangle$$

$\omega$ is the sixth root of unity and normalization factor is omitted for clarity. Note that these two bases are MU (Mutual Unbiased) [29], i.e. message encoded in one basis is completely random when measured by the other basis. This means that the message encoded in one base is completely hidden and lost permanently when measured by another MU base.

Now if the dealer wants to revoke the secret before a predefined date or event, he can ask t or more member to measure their quantum state $|x\rangle_\theta$ in the Fourier basis. He can verify the honesty of these members easily with his knowledge $x$ and $\theta$. If these t or members are honesty, he can also be assured that the secret is revoked. If any party is dishonest, the dealer can also easily identify them. But note that this simple extension lost the property of collaboration encouraging.

## 4. Security Analysis and Comparison

In this section, we show that our first CE-QSS-Seal construction is unconditional secure and matching with the optimal bound of a seal scheme [25]. Then we show our second CE-QSS-Seal construction exceed the optimal bound of [25], which is not a contradiction, since our second construction is not unconditionally secure but post-quantum secure.

### 4.1 Unconditional security of our first construction

**Theorem 1.** Our first CE-QSS-Seal construction above is unconditionally secure.

**Proof:** In our first construction, if one party is missing when reconstructing the secret, without loss of generality suppose the last party is missing, all the remaining $n-1$ parties put their qubits together will get a mixed state (for simplicity we omit the normalization factor):

$$|b_1 b_2 .. b_{n-1}\rangle\langle b_1 b_2 .. b_{n-1}| + |\overline{b_1}\overline{b_2}...\overline{b_{n-1}}\rangle\langle\overline{b_1}\overline{b_2}...\overline{b_{n-1}}|$$

If measured by computational basis, $b_1 b_2 .. b_{n-1}$ or $\overline{b_1}\overline{b_2}...\overline{b_{n-1}}$ will be gotten with equal probability. The remaining secret bit $b_n$ is completely unknown to the $n-1$ parties. So, to get the secret, the $n-1$ parties have to guess the final bit. Of course, the successful guessing probability is $1/2$.

For example, suppose ten parties share the 10-qubit quantum state:

$$|0011100101\rangle + |1100011010\rangle$$

If the first nine parties together measure their qubit by the computational basis, then they will get classical 9-bit long number 001110010 or 110001101 with equal probability. While the final bit is completely unknown to them. Hence they have to guess the final bit, the right guess would be 0011100101 or 1100011010, while 0011100100 and 1100011011 would be wrong guess. Clearly, the successful probability is $1/2$.

Similarly, if two parties are missing, then the successful guessing probability is $1/4$. And three missing parties correspond to $1/8$. In general, if there are k parties missing from the reconstruction process, then the successful recovering probability is $1/2^k$. I.e., the successful probability is diminishing exponentially with the number of missing parties. Hence if there are many missing parties, it is impossible to reconstruct the secret practically. Therefore, collaboration among participants is encouraged.

For the seal security, if some parties are dishonest, and try to get the secret before the predefined date or event. They have to measure the quantum state
$$|x\rangle + |\bar{x}\rangle$$
by the computational basis. These measurements will collapse the whole quantum state to $|x\rangle$ or $|\bar{x}\rangle$. So, if the dealer asks all the participant to send back their particles, the dealer will get $|x\rangle$ or $|\bar{x}\rangle$ instead of the original superposition state $|x\rangle + |\bar{x}\rangle$. When this quantum state is measured by the basis $\{|x\rangle + |\bar{x}\rangle, |x\rangle - |\bar{x}\rangle\}$, he will get $|x\rangle + |\bar{x}\rangle$ or $|x\rangle - |\bar{x}\rangle$ with equal probability. If the dealer gets $|x\rangle - |\bar{x}\rangle$, he knows that some of the participants are cheating. Hence with probability of $1/2$, the cheating behavior of the participants will be caught. Actually, this probability is optimal, since it matched with the upper bound in [25].

**4.2 Post-quantum security of our second scheme**

**Theorem 2.** Our second CE-QSS-Seal construction above is post-quantum secure with perfect reconstruction probability and cheat-detecting probability.

**Proof:** The proof of collaboration encouraging property is similar with above. Since each party holding one bit of the secret. If one party is missing, then the remaining $n-1$ parties should guess the bit and the successful guessing probability is $1/2$. And in general, if $k$ parties are missing, the successful guessing probability is $1/2^k$, same as above. So, more parties are involved, higher successful probability will be achieved. Hence collaborations among participants are encouraged.

For the seal property, if some parties are dishonest and try to reconstruct the secret before the predefined date or event. They have to decrypt the post-quantum secure ciphertext to get $\theta$ and $b \oplus (\oplus_{i:\, \theta_i=0} x_i)$, then measure the quantum $|x\rangle_\theta$ in the computational basis, recover the $x_i$s for which the corresponding $\theta_i$ is 0. Then unmask $b \oplus (\oplus_{i:\, \theta_i=0} x_i)\ldots$ to get the secret bit $b_i$. Notice that the quantum measurement is irreversible. Once it is carried out, the quantum state will collapse and cannot be reversed to the original state. Hence if the dealer asks them to return the measurement result by the Hadamard basis, they have no way to correctly come out those $x_i$s for which the corresponding $\theta_i$ is 1, i.e. those that encoded by the Hadamard basis. Hence with almost 100% probability the behavior will be caught by the dealer. More precisely, if half of the bit are encoded by the Hadamard basis, then the probability is $1 - 1/2^{n/2}$. This exceeds the optimal bound of [25], but not a contradiction. Since now it is not unconditional secure, only post-quantum computational secure.

Table 1. Comparison with optimal bound of quantum seal

|  | Security | Reconstruction probability | Cheat detecting Probability |
|---|---|---|---|
| Optimal Bound [25] | Unconditional | 100% | 50% |
| Ours 1 | Unconditional | 100% | 50% |
| Ours 2 | Post-Quantum | 100% | 100% |

In Table 1, we briefly compare our two constructions with the optimal bound of quantum seal in [25]. We can see that our first construction match with the optimal bound in [25]. I.e., if the successful reconstruction of the secret is 100%, then the cheating behavior can only be detected with 50% at most. And our second construction exceed the optimal bound, achieving 100% probability of cheat detecting when the successful reconstruction probability is 100%. But our second construction is not unconditionally secure, but post-quantum secure.

## 5. Conclusions

In this paper, we present a new concept of collaboration-encouraging quantum secret sharing with seal functionality. In CE-QSS-Seal, the secret can be easily reconstructed when all the parties are present. And the secret can also be reconstructed when few members are missing, but with more computational overhead. While the secret cannot be reconstructed in practice if many parties are missing, since the successful reconstruction probability diminishing exponentially with the number of missing parties.

The seal property of our scheme means that the dealer can revoke the secret before a predefined date or event by asking the participating parties return their shares. This is simply impossible in classical setting, since classical information can be easily copied any number of times. We show how to realize it quantum mechanically. First, we construct an unconditionally secure CE-QSS-Seal scheme matching the optional bound of quantum seal scheme. Then we continue to show how to improve this bound by post-quantum cryptography. I.e., we construct an almost perfect CE-QSS-Seal scheme based on public key encryption with deletion, which can be constructed based on post-quantum secure assumptions.

In the future, we plan to continue to research on how to extend our scheme to threshold secret sharing scenario, and how to improve the efficiency and security of our CE-QSS-Seal scheme.

**Acknowledgements**
This work is supported by Social Science Foundation of Fujian Province, China (FJ2024B088, FJ2024MGCA028).